\documentclass[aps,twocolumn,amsmath,amssymb,floatfix]{revtex4-1}
\usepackage{graphicx}
\usepackage{dcolumn}
\usepackage{bm}
\usepackage{amsfonts}
\usepackage{amsmath}
\usepackage{graphicx}
\usepackage{dcolumn}
\usepackage{bm}
\usepackage{amsmath}
\usepackage{amssymb}
\usepackage{verbatim}
\usepackage{mathtools}
\usepackage{epstopdf}

\begin{document}
\title{Diluted magnetic Dirac-Weyl materials:  Susceptibility and ferromagnetism in three-dimensional chiral gapless semimetals}
\author{Sanghyun Park$^{1}$}
\author{Hongki Min$^{1}$}
\email{Electronic address: hmin@snu.ac.kr}
\author{E. H. Hwang$^{2}$}
\email{Electronic address: euyheon@skku.edu}
\author{S. Das Sarma$^3$}
\affiliation{$^1$ Department of Physics and Astronomy, Seoul National University, Seoul 08826, Korea}
\affiliation{$^2$ SKKU Advanced Institute of Nanotechnology and Department of Nano Engineering, Sungkyunkwan University, Suwon, 16419, Korea}
\affiliation{$^3$Condensed Matter Theory Center and Joint Quantum Institute, Department of Physics, University of Maryland, College Park, Maryland  20742-4111}
\date{\today}

\begin{abstract}
We theoretically investigate the temperature-dependent static susceptibility and long-range magnetic coupling of three-dimensional (3D) chiral gapless electron-hole systems (semimetals) with arbitrary band dispersion [i.e., $\varepsilon(k) \sim k^N$, where $k$ is the wave vector and $N$ is a positive integer]. We study the magnetic properties of these systems in the presence of dilute random magnetic impurities. Assuming carrier-mediated Ruderman-Kittel-Kasuya-Yosida indirect exchange interaction, we find that the magnetic ordering of intrinsic 3D chiral semimetals in the presence of dilute magnetic impurities is {\it ferromagnetic} for all values of $N$. Using finite-temperature self-consistent field approximation, we calculate the ferromagnetic transition temperature ($T_{\rm c}$). We find that $T_{\rm c}$ increases with increasing $N$ due to the enhanced density of states, and the calculated $T_{\rm c}$ is experimentally accessible assuming reasonable coupling between the magnetic impurities and itinerant carriers.

\end{abstract}

\maketitle

\section{Introduction}
In recent years, there has been substantial interest in three-dimensional (3D) Weyl/Dirac semimetals, which have relativistic linear energy dispersion
\cite{Armitage2018,Burkov2015,Bansil2016,Wan2011, Burkov2011a, Burkov2011b, Hosur2012}.
These systems are effectively 3D versions of graphene which is the quintessential 2D Dirac system. 
The magnetic properties of Dirac-Weyl semimetals have been studied theoretically, demonstrating the possibility of magnetic ordering of the dopant magnetic impurities at zero temperature with and without spin-orbit coupling \cite{Chang2015,Hosseini2015,Sun2017}.  Mechanisms for various magnetic ordering in topological materials have been investigated, demonstrating that magnetically doped semiconductors with the strong spin-orbit interaction can have ferromagnetic ordering through the mechanism of van Vleck paramagnetism \cite{Yu2010, Kurebayashi2014}.
The spin susceptibilities in Weyl and Dirac semimetals have been calculated, investigating the effect of the magnetic texture and associated physical properties \cite{Araki2016, Zhou2018, Thakur2018, Ominato2018}.

The indirect exchange interactions between magnetic impurities through carriers of a host material (i.e., Ruderman-Kittel-Kasuya-Yosida (RKKY) interaction \cite{Ruderman1954,Kasuya1956,Yosida1957,Kittel1969}) in semimetal systems has become an interesting issue. 
Since the low energy dispersion in condensed matter systems can be nonlinear as in 2D multilayer graphene \cite{Min2008a,Min2008b} and 3D multi-Weyl semimetals \cite{Fang2012}, it is also interesting to find the effects of arbitrary band dispersion and finite temperature on the magnetic properties of 3D gapless systems in the presence of random magnetic impurities (i.e. in addition to the expected linear gapless chiral dispersion of Dirac systems).
Moreover, the nonlinear energy dispersions of itinerant carriers result in an interesting behavior of the RKKY interaction between magnetic spins, which provides a more complete picture of the qualitative nature of magnetic properties in gapless semimetals with arbitrary band dispersion.
In particular, it is useful to know whether 3D Dirac-Weyl gapless semimetals could magnetically order at finite temperatures through the RKKY coupling, and how the resultant magnetic transition temperature depends on the band carrier energy dispersion.

In this paper, we study the magnetic properties of 3D gapless electron-hole systems at finite temperatures with arbitrary band dispersion, focusing on the possibility of long-range ordering in the
magnetic moments that are embedded in the system. 
To study the carrier-mediated indirect RKKY exchange interaction among the random magnetic impurities with the itinerant carriers mediating the magnetic interaction between the impurities, we calculate the temperature-dependent response functions and the corresponding long-range magnetic coupling between dilute random magnetic impurities. We mainly focus on the dilute impurity limit, which is different from the strong disorder limit in which strong enough disorder may induce a phase transition to a metallic state \cite{Kobayashi2014, Sbierski2014,Pixley2016a,Pixley2016b}.
The effects of finite temperature, disorder, and carrier mean-free path on the RKKY interaction are also considered systematically in our current work, but we do not consider the topic of disorder-induced quantum phase transition since our focus is on magnetic properties and not disorder physics. Especially, we study the role of the ultraviolet momentum cutoff, which is necessary in gapless semimetals, demonstrating that it fundamentally modifies the characteristic power-law behavior of the RKKY interaction. Inclusion of the mean-free path in the theory allows us to make a specific prediction about the dependence of the magnetic behavior of the system on the carrier transport properties \cite{DasSarma2015}.
A smooth interpolation between long-range and short-range magnetic interactions is possible by varying the cutoff parameter $R$ in the range of the RKKY interaction, which is related to the localization length of the carriers in semimetals \cite{Priour2004,Priour2005, Min2017}.

By considering all these effects together within one comprehensive mean-field theory, 
we calculate the ferromagnetic transition temperature in the framework of a finite-temperature self-consistent field approximation \cite{Priour2004} 
for the ferromagnetism in 3D gapless semimetals.
We find that in 3D gapless semimetals, the ferromagnetic ordering between magnetic impurities induced by the RKKY exchange interaction is favored with enhanced magnetic coupling, as the energy dispersion has a higher power-law. Our results indicate that within the experimentally accessible range of parameters, ferromagnetic ordering between magnetic impurities is possible in gapless 3D semimetals. Ferromagnetism in 3D semimetals, as predicted in our theory, can be utilized in spintronics applications if our predictions are validated experimentally.
We predict that it should be possible to experimentally induce long-range finite-temperature ferromagnetic ordering in 3D Dirac-Weyl materials by magnetically doping the system.

This paper is organized as follows. In Sec.~\ref{sec:model}, we describe our model and calculate the finite-temperature static susceptibilities. In Sec.~\ref{sec:RKKY_interaction}, we provide the calculated results of the effective magnetic coupling through RKKY interaction in 3D chiral gapless semimetals. The conclusions are provided in Sec.~\ref{sec:discussion_and_conclusion} with a discussion on
the momentum-cutoff effect on long-range oscillations.

\section{Model}
\label{sec:model}

To describe the 3D chiral gapless semimetals, including Weyl/Dirac semimetals, we introduce the following Hamiltonian with an isotropic energy dispersion characterized by a positive integer $N$ \cite{Ahn2016}:
\begin{equation}\label{eq:3d_chiral_hamiltonian}
H=\varepsilon_0 \left(\frac{|\bm{k}|}{k_0} \right)^N \hat{\bm{k}} \cdot \bm{\sigma},
\end{equation}
where $\bm{\sigma}$ represents the Pauli matrices acting in the space of the two bands near the band touching point, and $\varepsilon_0$ and $k_0$ are materials dependent constants with dimensions of energy and wave vector, respectively. 
We note that the introduced Hamiltonian describes a gapless electron-hole system with arbitrary energy-band dispersion. Even though the real systems with the Hamiltonian except a linear dispersion ($N=1$) may not be available currently in 3D, it is interesting to obtain the magnetic properties for both linear and nonlinear energy dispersions to develop intuition about the dispersion dependence of RKKY interactions.
Also, rapid development in the materials science may lead to such materials with nonlinear dispersion in the future, making our theory for the nonlinear dispersion of experimental relevance.
Here, for simplicity, we assume that the Pauli matrices describe the pseudospin degrees of freedom rather than real spin degrees of freedom, focusing on the effect of the arbitrary energy
dispersion. We will discuss the effect of the real spin texture in the Discussion and Conclusion section. The energy dispersion of the Hamiltonian is given by $\varepsilon_{\lambda,\bm{k}} =\lambda \varepsilon_0 \left(\frac{|\bm{k}|}{k_0} \right)^N$, where $\lambda= \pm1$ is the band index for the  conduction (valence) band. Note that the Hamiltonian with $N=1$ corresponds to Weyl semimetals with linear energy dispersion. We assume that the system is intrinsic with the Fermi energy at the band touching point, which we take to be the zero of energy.
We are thus considering undoped intrinsic Dirac-Weyl systems with the chemical potential pinned at the Dirac-Weyl point.


Carrier-mediated RKKY indirect exchange interaction between local moments is proportional to the static carrier susceptibility. At finite temperatures, the static susceptibility is given by
\begin{equation}
\label{eq:static_polarizability}
\chi(\bm{q},T)=-g\sum_{\lambda,\lambda'} \int{\frac{d^3\bm{k}}{(2\pi)^3}\frac{f_{\lambda,\bm{k}}-f_{\lambda',\bm{k}'}}{\varepsilon_{\lambda,\bm{k}}-\varepsilon_{\lambda',\bm{k}'}}}F_{\lambda,\lambda'}(\bm{k},\bm{k}'),
\end{equation}
where $g$ is the total (e.g., spin, valley, etc.)
degeneracy factor, $f_{\lambda,\bm{k}}= \left[ e^{{\varepsilon_{\lambda,\bm{k}}}/{k_{\rm{B}}T}}+1\right]^{-1}$ is the finite-temperature Fermi-Dirac distribution function for the $\lambda$-band and wave vector $\bm{k}$, the chiral factor
$F_{\lambda,\lambda'}(\bm{k},\bm{k}')$ is the square of the wavefunction overlap between $\left|\lambda,\bm{k}\right>$ and $\left|\lambda',\bm{k}'\right>$ states, and $\bm{k}'=\bm{k}+\bm{q}$. 
For the 3D chiral gapless system described by Eq.~(\ref{eq:3d_chiral_hamiltonian}), $F_{\lambda,\lambda'}(\bm{k},\bm{k}') = \frac{1}{2}\left(1+\lambda \lambda' \cos \theta_{\bm{k},\bm{k}'} \right)$ for all $N$, where $\theta_{\bm{k},\bm{k}'}$ is the angle between $\bm{k}$ and $\bm{k}'$. 
Note that $F$ arises entirely from the chirality of the system.

Dividing the sum in Eq.~(\ref{eq:static_polarizability}) into interband ($\lambda \neq \lambda'$) and intraband ($\lambda = \lambda'$) contributions, the static susceptibility can be decomposed into $
\chi(\bm{q},T)= \chi^{+}(\bm{q},T)+\chi^{-}(\bm{q},T)$, where $\chi^{\pm}$ denote the interband ($+$) and intraband ($-$) contributions, respectively.  
With the density of states (DOS) at $T=0$, $D_N(q)= \frac{gq^{3-N}k_0^N}{2\pi^2 N \varepsilon_0}$, the normalized susceptibility $\chi^{\pm}$ can be rewritten as
\begin{eqnarray}
\label{eq:normalized_static_polarizability}
{\chi^{\pm}(\bm{q},T)\over D_N(q)}
&=& {N\over 4} \int_{0}^{\infty}x^2 dx\int_{0}^{\pi}{\sin\theta d\theta \frac{\left(1\mp \cos\psi \right)}{x^N\pm (x')^N}} \nonumber \\
&\times& \left[\tanh \frac{(qx/k_0)^N}{2T/T_0} \pm \tanh \frac{(qx'/k_0)^N}{2T/T_0} \right],
\end{eqnarray}
where $\theta$ is the angle between $\bm{k}$ and $\bm{q}$, and $\psi$ is the angle between $\bm{k}$ and $\bm{k}'=\bm{k}+\bm{q}$. In Eq.~(\ref{eq:normalized_static_polarizability}), $x=k/q$, $x'=k'/q=\sqrt{1+2x\cos\theta+x^2}$, $\cos\psi=(x+\cos\theta)/x'$, and $T_0 = \varepsilon_0/k_{\rm B}$. Note that for $N=1$, a finite (ultraviolet) momentum cutoff is required for the convergence of the integral. For the calculation, we set $g=4$ and $k_0=a^{-1}$, where $a$ is the lattice constant of the system.

At zero temperature ($T=0$), due to the phase-space restriction, the intraband part $\chi^{-}$ vanishes and only the interband part $\chi^{+}$ contributes to the total susceptibility. In the long wavelength limit ($q\rightarrow 0$), the susceptibility approaches the DOS and  $\chi^{+}(q,T=0)\propto q^{3-N}$, which diverges for $N\ge 4$ as $q\rightarrow 0$. At finite temperatures ($T\neq 0$), we obtain $\chi^{+}(q,T)\propto q^3/T$ for $q \rightarrow 0$. Thus, the $q=0$ singularity of $\chi^{+}(q,T=0)$ for $N\ge 4$ disappears. In addition, due to the thermal excitation of electrons and holes, $\chi^{-}$ also contributes to the susceptibility at finite temperatures even for the undoped system under consideration.
Thus, for $T \neq 0$, the total susceptibility at $q=0$ becomes finite for all $N$.
Specifically, we find that $\chi^{-}(q\rightarrow 0,T)\propto T^{3-N \over N}$, which shows the same power-law dependence as the DOS, $D_N(\varepsilon)\propto \varepsilon^{3-N \over N}$ with energy $\varepsilon$ replaced by $T$. Therefore, as temperature increases, the total susceptibility at $q=0$ increases for $N=1,2$, remains constant for $N=3$, and decreases for $N\ge 4$.
These analytical findings are helpful in understanding our detailed numerical results presented in the rest of this paper.

\begin{figure}[t]
\vspace{10pt}%
\includegraphics[width=0.8\linewidth]{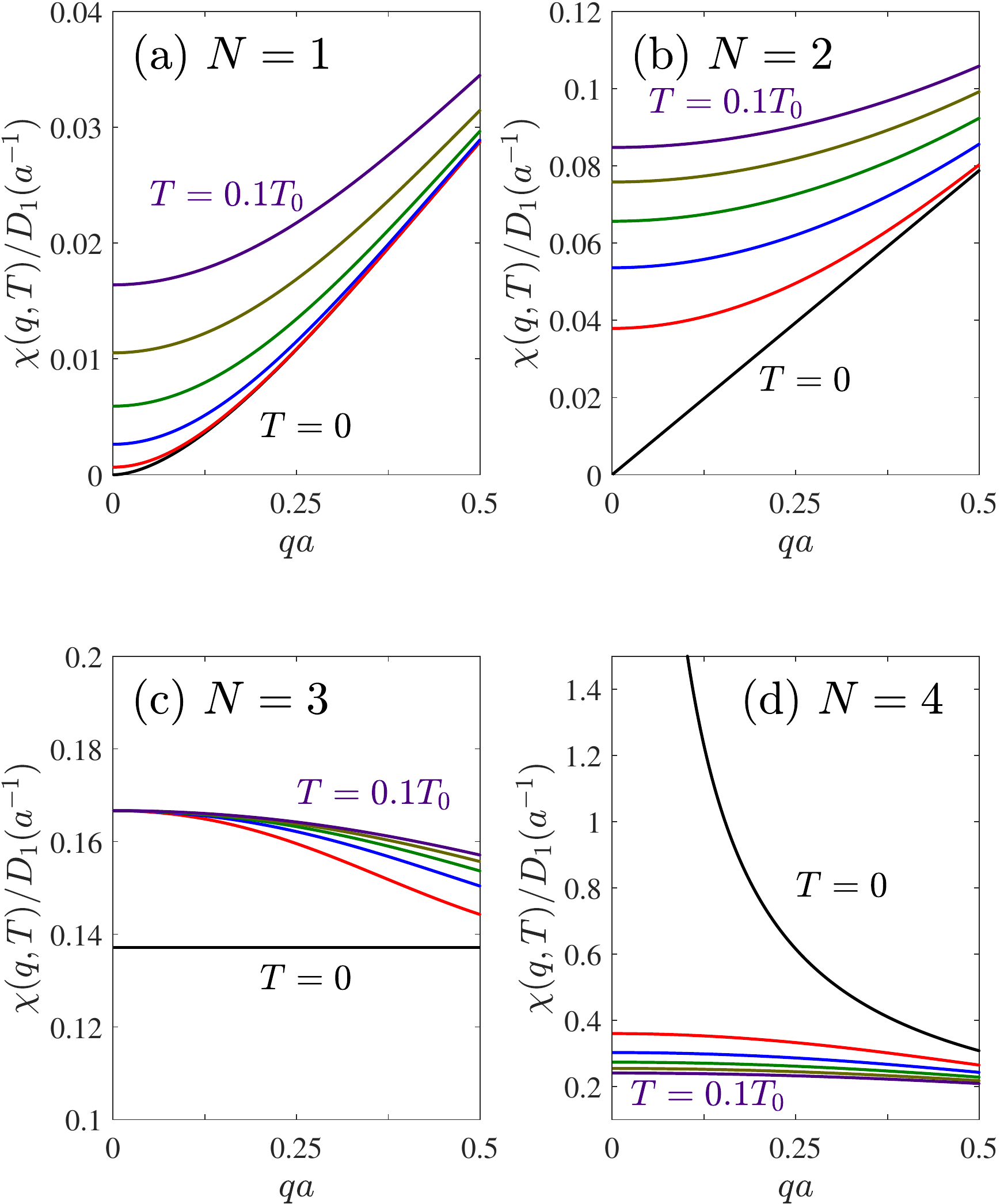}
\caption{
The calculated finite-temperature static susceptibility $\chi(\bm{q},T)$ as a function of wave vector for various temperatures $T=0, 0.02, 0.04, 0.06, 0.08,$ and $0.1$ $T_0$, and for different values of $N$ (a) $N=1$, (b) $N=2$, (c) $N=3$, and (d) $N=4$. Here, $T_0=\varepsilon_0/k_{\rm B}$, $D_1(a^{-1})={g k_0 \over 2\pi^2 \varepsilon_0 a^2}$, and $a=0.343$ nm (lattice constant of TaAs). 
For $N=1$, the finite momentum cutoff $a^{-1}$ is used for the convergence of the integral.
}
\label{fig:polarization_function}
\end{figure}

Figure~\ref{fig:polarization_function} shows the calculated static susceptibility as a function of the wave vector for several temperatures. For $N=1,2$, the susceptibility increases with temperature, whereas for $N\ge 4$, it decreases with temperature, as expected. 
Interestingly, for $N\ge 3$, the finite-temperature result in the $T \rightarrow 0$ limit is different from the zero-temperature value, i.e., $\chi(0,T=0)\neq \chi(0,T\rightarrow 0)$. 
Note that for $N=3$, the DOS $D_N(q)$ becomes constant and $\chi(0,T\rightarrow 0)$ approaches the constant DOS, whereas $\chi(0,T=0)$ can be obtained from Eq.~(\ref{eq:normalized_static_polarizability}). 
For $N=3$, we find that $\chi(0,T=0)/\chi(0,T\rightarrow 0)\approx0.8229$. 
This $T=0$ non-analyticity in the $N=3$ susceptibility follows from the fact that the DOS has a `kink' structure at $N=3$ with $D_N(\varepsilon)$ increasing (decreasing) as a function of increasing energy for $N<3$ $(N>3)$.


\section{RKKY interaction and effective magnetic coupling}
\label{sec:RKKY_interaction}


To study the effective magnetic coupling between random magnetic impurities
(which are treated as quenched classical magnetic moments),
we consider the carrier-mediated RKKY indirect exchange interaction. The indirect exchange interaction between magnetic impurities can be accounted for by the interaction between a localized (classical)
spin $\bm{S}_i$ of a magnetic impurity located at $\bm{r}_i$ and an itinerant electron spin $\bm{s}$ located at $\bm{r}$. It is  given by $V(\bm{r})=J_{\rm{ex}}\bm{S}_i \cdot \bm{s} \delta(\bm{r}_i-\bm{r})$, where $J_{\rm{ex}}$ is the local exchange coupling
between the quenched impurity and the itinerant carriers.  ($J_{\rm ex}$, which depends on the nature of the magnetic impurities, is an unknown parameter in our theory providing the overall magnitude of the magnetic coupling in the system.)
Then, the effective Hamiltonian that describes the magnetic interactions between the classical Heisenberg spins $\bm{S}_{i}$ and $\bm{S}_j$ located at $\bm{r}_{i}$ and $\bm{r}_j$, respectively, is given by 
\begin{equation}
H=-\sum_{i,j}J_{\rm{RKKY}}(\bm{r}_i-\bm{r}_j)\bm{S}_i \cdot \bm{S}_j,
\end{equation}
where
\begin{equation}
J_{\rm{RKKY}}(\bm{r},T)=\frac{[J_{\rm{ex}}a^3]^2}{4}\chi(\bm{r},T).
\end{equation}
The RKKY range function $\chi(\bm{r},T)$ is defined by the Fourier transform of the static susceptibility $\chi(\bm{q},T)$. For an isotropic system in 3D, it is given by
\begin{equation}
\label{eq:range_function}
\chi(\bm{r},T)=\frac{1}{2\pi^2} \int_{0}^{\infty} {q^{2} dq} j_{0}(qr)\chi(\bm{q},T),
\end{equation}
where $ j_{0}(x)$ is the spherical Bessel function of the first kind. 
Since the large momentum cutoff $q_{\rm c}$ is natural for a continuum theory, we set $q_{\rm c}=a^{-1}$ in the numerical calculation of the range function in Eq.~(\ref{eq:range_function}). 


\begin{figure}[t]
\vspace{10pt}%
\includegraphics[width=0.8\linewidth]{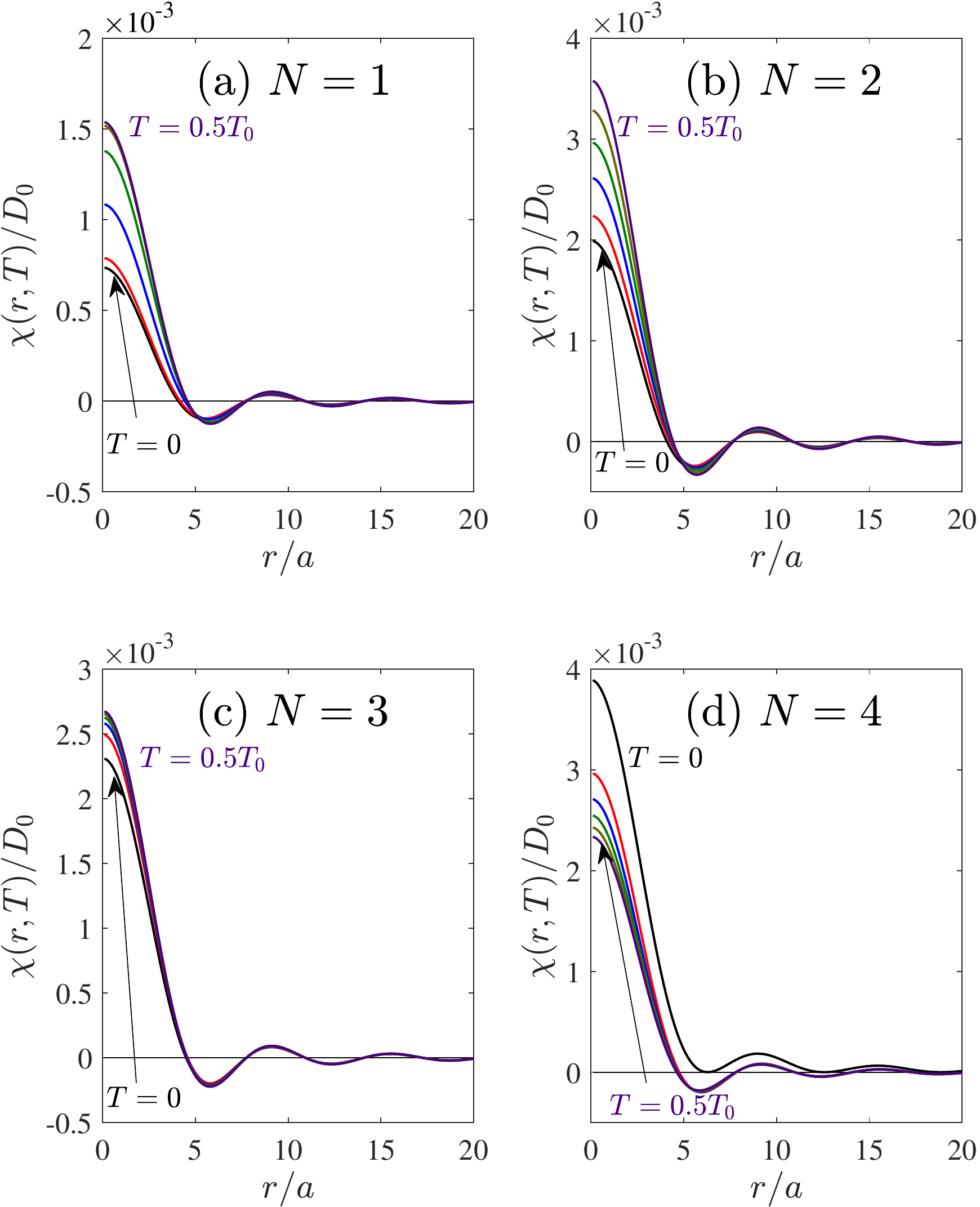}
\caption{
The range function $\chi(\bm{r},T)$ as a function of distance for different values of $N$ (a) $N=1$, (b) $N=2$, (c) $N=3$, and (d) $N=4$. In each figure, the curves with different colors represent different temperatures $T =$ 0, 0.1, 0.2, 0.3, 0.4, and 0.5 $T_0$. Here, $D_0=D_1(a^{-1})/a^3$. In this calculation, the ultraviolet momentum cutoff $q_{\rm c}=a^{-1}$ is used. 
}
\label{fig:range_function}
\end{figure}

Figure~\ref{fig:range_function} shows the range functions for $N=1,2,3,4$ and for different temperatures. 
For $N=1,2$, the magnitude of the oscillating range functions increases with temperature, whereas for $N\ge 4$, it decreases with temperature. For $N=3$, the range function is almost independent of temperature. These behaviors for different $N$ follow from the temperature dependence of the susceptibility, which is shown in Fig.~\ref{fig:polarization_function}. 
At large distances ($r/a\gg 1$), we find that the range function decays as $\cos(q_{\rm c} r)/r^2$ for $N\le 3$, producing long-range oscillations with a periodicity of $2\pi/q_{\rm c}$ in the spin density. (See Appendix~ \ref{sec:cutoff_dependence_on_the_decaying_behavior_of_the_range_function} for the detailed derivations.) We will discuss the implications of the cutoff $q_{\rm c}$ in the Discussion and Conclusion section (see Sec.~\ref{sec:discussion_and_conclusion}).

The temperature-dependent effective coupling is given by the spatial average of the RKKY interaction $J_{\rm{RKKY}}$,
\begin{equation}
\label{eq:effective_coupling}
J_{\rm{eff}}(T)=\frac{1}{\Omega_{\rm{unit}}}\int {d^3r J_{\rm{RKKY}}(\bm{r},T)},
\end{equation}
where $\Omega_{\rm{unit}}$ is the volume of a unit cell. In the dimensionless form, Eq.~(\ref{eq:effective_coupling}) can be rewritten as
\begin{equation}
{J_{\rm{eff}}(T) \over J_{\rm{eff}}^{(0)}}={1\over D_0(a^{-1})}\int {r^2 dr \chi (\bm{r},T)},
\end{equation}
where $J_{\rm{eff}}^{(0)}=4\pi [J_{\rm{ex}}a^3]^2 D_0(a^{-1}) /4\Omega_{\rm{unit}}$ and $D_0=D_1(a^{-1})/a^3$. Note that the normalization factors $J_{\rm{eff}}^{(0)}$ and $D_0$ are defined to be independent of both index $N$ and temperature $T$.

In the presence of non-magnetic impurity scattering arising from unintentional background disorder causing momentum relaxation, the
RKKY interaction should be cut off at distances larger than a characteristic disorder length scale
(i.e., the transport mean-free path),
which is determined by the impurity scattering. We include the disorder effect phenomenologically by including an exponential damping at distances larger than the cutoff $R$ in the range of the RKKY interaction. Then, the effective coupling is modified as
\begin{equation}
\label{eq:effective_coupling_exponential_damping}
J_{\rm{eff}}(T)=
\begin{cases}
\frac{1}{\Omega_{\rm{unit}}}\int{d^3r J_{\rm{RKKY}}(\bm{r})} & (r<R), \\
\frac{1}{\Omega_{\rm{unit}}}\int{d^3r J_{\rm{RKKY}}(\bm{r})}e^{-\frac{r-R}{R}} & (r>R).
\end{cases}
\end{equation}
(See Appendix~\ref{sec:effective_RKKY_coupling_with_the_exponential_cutoff} for the detailed expression of the effective RKKY coupling with exponential cutoff.)
In this calculation, we use $R=100 a$, and our calculated results do not depend on the choice of $R$ qualitatively. 
One should think of $R$ as a disorder-induced phenomenological effective carrier mean-free path parameter, which provides a cutoff for the RKKY interaction range.  $R$ should in general be smaller (larger) depending on the system being more (less) disordered.  As a matter of principle, $R$ cannot really be very large since the magnetic ordering phenomenon being studied here necessitates the presence of magnetic impurities, which, in addition to providing the quenched magnetic moments for ordering, also serve as momentum scatterers.


\begin{figure}[t]
\vspace{10pt}%
\includegraphics[width=0.78\linewidth]{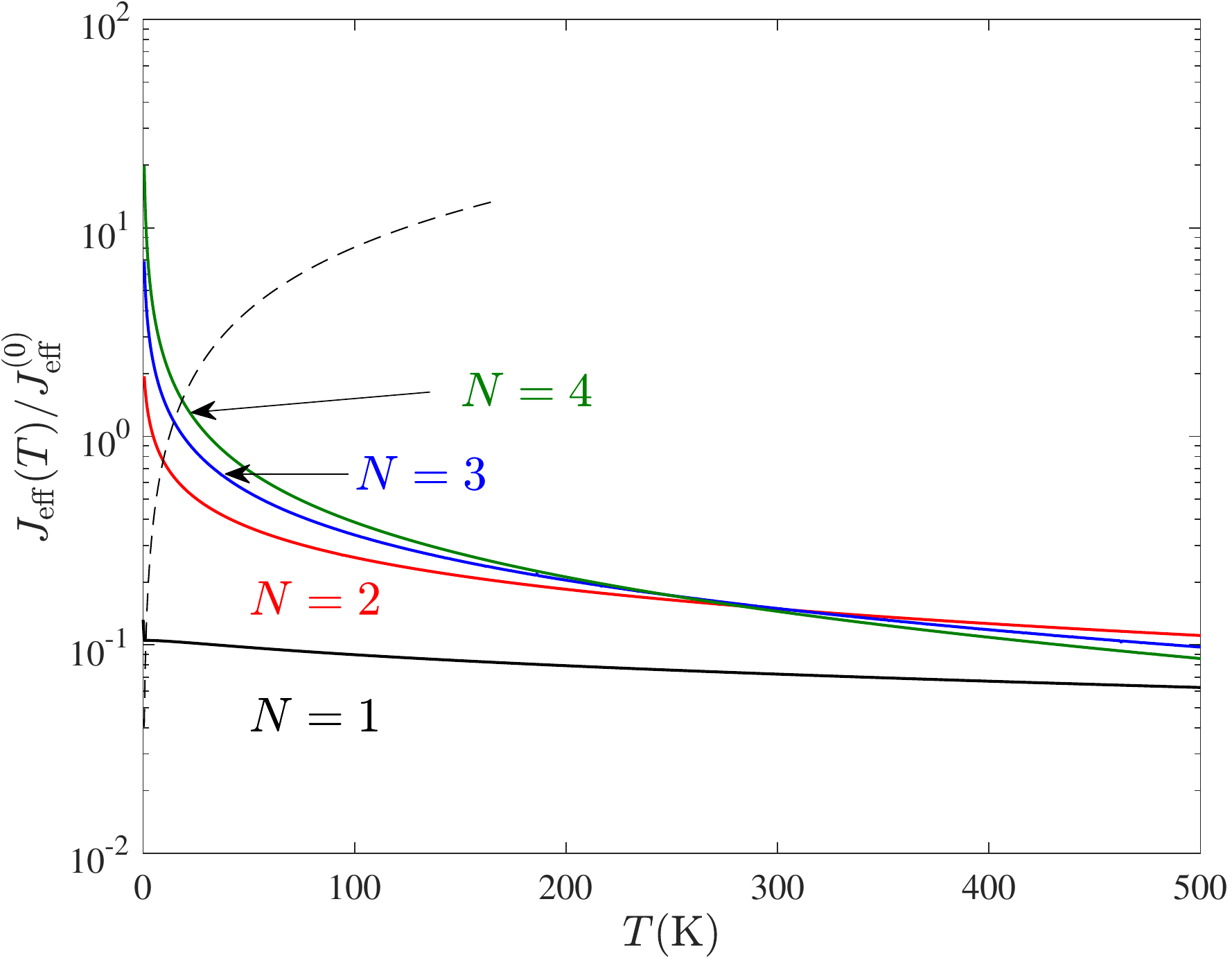}
\caption{
The calculated effective coupling (solid lines) as a function of temperature for different values of $N=1,2,3,4$. In this calculation, the ultraviolet cutoff $q_{\rm c}=a^{-1}$ and exponential cutoff $R=100a$ are used. Here, the normalization factor $J_{\rm{eff}}^{(0)}=4\pi [J_{\rm{ex}}a^3]^2 D_1(a^{-1})/4\Omega_{\rm{unit}}$ is independent of $N$ and temperature $T$. The dashed line represents $3k_{\rm{B}}T/[S(S+1)x]$, and the intersections with $J_{\rm{eff}}(T)$ indicate the transition temperatures solved self-consistently. Here, $J_{\rm{ex}}=0.1$ eV , $x=0.05$ and $S=5/2$.
}
\label{fig:effective_rkky}
\end{figure}


Figure~\ref{fig:effective_rkky} shows the calculated effective coupling as a function of temperature for different values of $N$. The effective coupling decreases monotonically with increasing temperature and increases with increasing $N$ at a fixed temperature. Since the effective coupling $J_{\rm eff}(T)$ is positive, the magnetic moments are expected to be ferromagnetically aligned.


From the temperature dependent effective coupling in Eq.~(\ref{eq:effective_coupling_exponential_damping}), we calculate
the magnetic transition temperature of the intrinsic chiral 3D semimetals.
For the Heisenberg classical spins, the mean-field transition temperature $T_{\rm c}$ is given by \cite{Kittel2005,DasSarma2003}
\begin{equation}
\label{eq:T_c}
k_{\rm{B}}T_{\rm c} =\frac{S(S+1)}{3} xJ_{\rm{eff}},
\end{equation}
where $S$ and $x=n_{\rm{imp}}a^3$ are the spin and concentration of the local magnetic moments, respectively, and $n_{\rm imp}$ is the effective magnetic impurity density. Since the calculated $J_{\rm eff}$ is a function of temperature, we calculate $T_{\rm c}$ self-consistently from Eq.~(\ref{eq:T_c}).
In Fig.~\ref{fig:effective_rkky}, the intersections between the dashed line [i.e., $3k_{\rm{B}}T/[S(S+1)x]$] and solid lines [i.e., $J_{\rm{eff}}(T)$] determine the transition temperature for each $N$.
 

\begin{figure}[t]
\vspace{10pt}%
\includegraphics[width=1\linewidth]{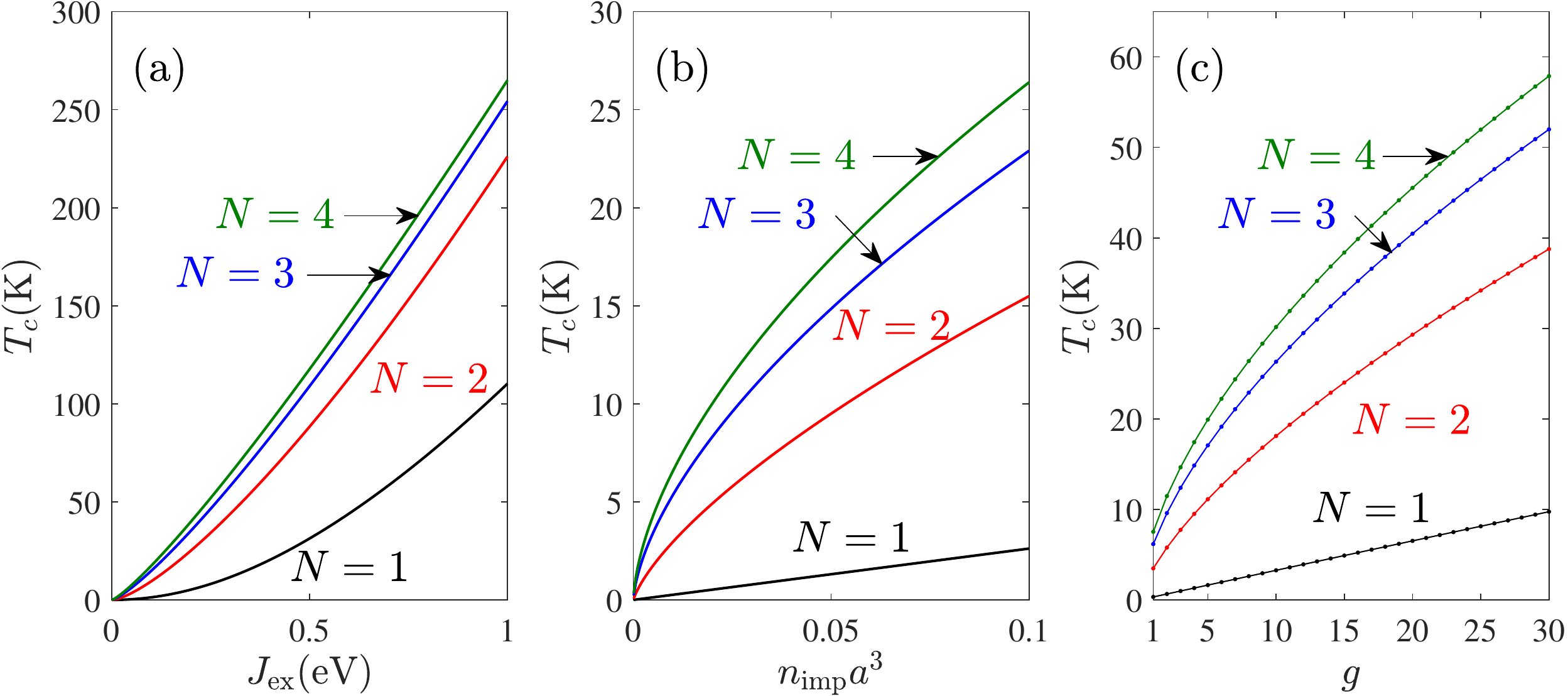}
\caption{
The calculated transition temperature $T_{\rm c}$ as a function of (a) the exchange coupling $J_{\rm{ex}}$ , (b) the magnetic impurity concentration $x=n_{\rm imp}a^3$, and (c) the degeneracy factor $g$ for different values of $N=1,2,3,4$. Here, for fixed parameters, we used $J_{\rm ex}=0.1$ eV, $x=0.05$ and $g=4$.
}
\label{fig:tc_jex_nimp}
\end{figure}


Figure ~\ref{fig:tc_jex_nimp} shows the self-consistently calculated transition temperature for different values of $N$ as a function of the exchange coupling $J_{\rm ex}$, the magnetic impurity concentration $x$, and the degeneracy factor $g$. The ferromagnetic transition temperature increases monotonically with increasing $J_{\rm ex}$, $x$, and $g$ for all $N$. In particular, for $N=1$, $T_{\rm c}$ increases quadratically with $J_{\rm ex}$, and linearly with both $x$ and $g$, as expected \cite{Kittel2005, DasSarma2003}. However, for $N>1$, the calculated $T_{\rm c}$ shows non-trivial dependence on the parameters arising from self-consistency even for small values of the parameters, due to the non-trivial behavior of the temperature-dependent effective coupling shown in Fig.~\ref{fig:effective_rkky}.

\section{Discussion and conclusion}
\label{sec:discussion_and_conclusion}


We studied theoretically the effective magnetic coupling between magnetic impurities and the 
consequent ferromagnetic transition temperature in 3D chiral gapless semimetals with arbitrary energy dispersion. To calculate the RKKY magnetic coupling
range function, we introduced the momentum cutoff $q_{\rm c}$, which is natural for the effective continuum model used in this work. 
We use the inverse lattice constant as the natural ultraviolet momentum cutoff in the theory.
As shown in Fig.~\ref{fig:range_function}, we find that for $N\le 3$, the envelope of the oscillatory RKKY range function decays as $r^{-2}$ and the period of the oscillation is $2\pi/q_{\rm c}$.
This decaying pattern arises from the finite ultraviolet cutoff $q_{\rm c}$ used in the range function. If we set the cutoff to be infinite, then the range function loses its oscillatory characteristics and monotonically decays as $r^{-6+N}$, which gives the typical $r^{-5}$ decay for $N=1$ \cite{Chang2015,Hosseini2015}.
However, in the presence of a finite $q_{\rm c}$, the overall behavior of the range function is determined by the competition between the oscillatory $r^{-2}$ term and the non-oscillatory $r^{-6+N}$ term. We find that for $N \le 3$, the oscillatory $r^{-2}$ decay dominates over the non-oscillatory $r^{-6+N}$ term, but for $N>3$ it is vice versa (see Appendix~\ref{sec:cutoff_dependence_on_the_decaying_behavior_of_the_range_function} for the detailed derivations). We note that the cutoff dependence of the effective coupling and the corresponding transition temperature is insensitive to the precise quantitative choice of $q_{\rm c}$. If the Pauli matrices in Eq.~(\ref{eq:3d_chiral_hamiltonian}) refer to real spin degrees of freedom, in addition to the Heisenberg-type spin-spin interaction term, there appears the Ising-type spin-spin interaction term \cite{Chang2015,Hosseini2015}. The systematic evaluation of the transition temperature in that situation is beyond the scope of the current paper, but we expect that the same power-law dependence which is fundamentally affected by presence of the cutoff will appear in this case. 
The study of the RKKY physics in the presence of both Heisenberg and Ising couplings remains an interesting theoretical problem for the future.

In summary, we investigate the temperature dependent susceptibility, RKKY interaction, and effective magnetic ordering for 3D chiral gapless semimetals with arbitrary energy dispersion
in the presence of dilute random magnetic impurities.
We find that in 3D chiral gapless semimetals, the ferromagnetic ordering between magnetic impurities is favored with enhanced magnetic coupling as the energy dispersion has a higher power-law.
Our results indicate that ferromagnetic ordering between magnetic impurities is possible in 3D gapless semimetals,  arising entirely
from the carrier-mediated indirect RKKY interaction in the dilute impurity limit. This predicted ferromagnetic ordering between magnetic impurities should be experimentally accessible with suitable magnetic doping.
Our theory is valid when quantum fluctuations and direct exchange coupling between the impurity moments are negligible, which should be justified for large impurity spins and dilute impurity concentrations.
In this paper, we consider only the case of zero Fermi energy, and the effect of finite Fermi energy would be an interesting future research direction.
Our finding that even the intrinsic undoped semimetallic system could be converted to a ferromagnet by dilute magnetic doping has obvious experimental implications, which should be explored in the laboratory.

\acknowledgments

This work was supported by the National Research Foundation of Korea (NRF) funded by the Korea government (MSIT) (Grant No.~2018R1A2B6007837) and Creative-Pioneering Researchers Program through Seoul National University (S.P. and H.M.), NRF-2017R1A2A2A05001403 (E.H.H), and the Laboratory for Physical Sciences (S.D.S).

\appendix

\section{Cutoff dependence of the range function}
\label{sec:cutoff_dependence_on_the_decaying_behavior_of_the_range_function}

To derive the asymptotic behavior of Eq.~(\ref{eq:range_function}) at zero temperature, we use the fact that the zero-temperature intrinsic polarization function $\chi(\bm{q},T=0)$ is proportional to the DOS, which is given by
\begin{equation}\label{eq:approx_pol}
\chi(\bm{q},T=0) \propto q^{d-N},
\end{equation}
where $d$ is the dimension of the system. Then, the range function $\chi(\bm{r}) \equiv \chi(\bm{r}, T=0)$ becomes
\begin{equation}\label{eq:approx_range_function}
\chi(\bm{r}) = C \int_0^{q_{\rm c}}dq q^{2d-N-1} f_d(qr),
\end{equation}
where $C$ is a momentum- and position-independent constant. Here, $f_d(qr)$ is defined by
\begin{equation}
f_d(qr)=\begin{cases}
J_0(qr) & (d=2), \\
j_0(qr) & (d=3),
\end{cases}
\end{equation}
where $J_0(x)$ and $j_0(x)={\sin x \over x}$ are the Bessel and spherical Bessel functions of the first kind, respectively. 

First, consider the 3D case. Using the following integral (see Eq.~(3.761) in Ref.~\cite{Gradshteyn2015}) 
\begin{equation}
\int_0^1 dx x^{\mu-1}\sin (ax)= \frac{-i}{2\mu}\left[M(\mu,\mu+1,ia)-{\rm c.c.}\right] ,
\end{equation}
where $a > 0$,  ${\rm{Re}}(\mu) >-1$, $\mu \neq 0$, and  $M(a,b,z)={}_{1}F_1(a;b;z)$ is the Kummer's confluent hypergeometric function, we obtain $\chi(\bm{r})$ as 
\begin{eqnarray}
\chi(\bm{r})\! =\!\frac{C}{r^{6-N}}\!\frac{-i (q_{\rm c} r)^{5-N}}{2(5-N)}\left[M(5-N,6-N,i q_{\rm c} r)-{\rm c.c.}\right]. \nonumber \\
\end{eqnarray}
The asymptotic behavior of $M(a,b,z)$ at a large $z$ is given by (see p.~508 in Ref.~\cite{AS1964})
\begin{eqnarray}
M(a,b,z)\approx \Gamma(b) \left[ \frac{e^z z^{a-b}}{\Gamma(a)} + \frac{(-z)^{-a}}{\Gamma(b-a)} \right]. 
\end{eqnarray}
Therefore, at large distances ($q_{\rm c} r\gg1$), $\chi(\bm{r})$ in 3D can be expressed as follows:
\begin{equation}
\chi(\bm{r})  \approx \frac{A\cos(q_{\rm c} r)}{r^2}+\frac{B}{r^{6-N}},
\label{a7}
\end{equation}
where $A$ and $B$ are constants.
When $6-N\le 2$, i.e., $N \ge 4 $, the second term in Eq.~(\ref{a7}) dominates over the first term. Since the magnitude of the oscillating term is smaller than that of the monotonic decaying term, the oscillation of the range function mostly occurs at positive values. 
In contrast, for $N\le 3$, the oscillating first term dominates. Therefore,
the range function oscillates with a period $2\pi/q_{\rm c}$, and its amplitude decays as $1/r^{2}$.

Similarly, for 2D, we find that when $4-N\le {3\over 2}$, i.e., $N=3,4,\cdots$, the range function decays as $1/r^{4-N}$, while for $N=1,2$, it oscillates with a period $2\pi/q_{\rm c}$, and its amplitude decays as $1/r^{\frac{3}{2}}$. This result is consistent with Min \textit{et al.} \cite{Min2017} 
for 2D gapless semimetals.

\section{Effective RKKY coupling with the exponential disorder cutoff}
\label{sec:effective_RKKY_coupling_with_the_exponential_cutoff}
From Eq.~(\ref{eq:effective_coupling_exponential_damping}) in the main text, the normalized effective RKKY coupling with exponential damping is given by
\begin{eqnarray}
\frac{J_{\rm{eff}}(T)}{J_{\rm{eff}}^{(0)}}&=& \int_0^{R/a} {\tilde{r}^2d\tilde{r} \tilde{\chi}(\bm{r},T) } \nonumber \\
&+&\left(\int_0^{\infty} -\int_0^{R/a}\right) {\tilde{r}^2d\tilde{r} \tilde{\chi} (\bm{r},T) e^{-\frac{r-R}{R}}} \\
&=&\int_0^{R/a} {\tilde{r}^2d\tilde{r} \tilde{\chi} (\bm{r},T)\left(1- e^{-\frac{r-R}{R}}\right)} + F(T) \nonumber,  
\end{eqnarray}
where $\tilde{r}=r/a$, $\tilde{\chi}(\bm{r},T)=\chi(\bm{r},T)/D_0$, and
\begin{eqnarray}\label{eq:f_t minimal}
F(T) &\equiv& \int_0^{\infty} {\tilde{r}^2d\tilde{r} \tilde{\chi} (\bm{r},T)e^{-\frac{r-R}{R}} },
\end{eqnarray}
which can be rewritten as (see Eq.~(6.623) in Ref.~\cite{Gradshteyn2015})
\begin{eqnarray}
F(T) &\equiv& \int_0^{\infty} {\tilde{r}^2d\tilde{r} \tilde{\chi} (\bm{r},T)e^{-\frac{r-R}{R}} } \\ \nonumber 
&=& \int_{0}^{q_{\rm c}a} {\frac{\tilde{q}^2d\tilde{q}}{2\sqrt{2\tilde{q}\pi^3}}} \frac{2e\left(2q\right)^{1 \over 2}\Gamma(2)}{R\sqrt{\pi}\left(1/R^2+q^2\right)^{2}} \tilde{\chi} (\bm{q},T)\\ \nonumber
&=& \int_{0}^{q_{\rm c}a} {\frac{\tilde{q}^2d\tilde{q}}{\pi^2}} \frac{e\tilde{R}^3 \tilde{\chi} (\bm{q},T)}{\left(1+\tilde{q}^2\tilde{R}^2\right)^{2}}, \nonumber
\end{eqnarray}
where $\tilde{q}=qa$, $\tilde{R}=R/a$, $\tilde{\chi} (\bm{q},T)=\chi (\bm{q},T)/D_1(a^{-1})$, and $\Gamma$ is the gamma function. 


\begin{thebibliography}{10}

\bibitem{Armitage2018}
N.~Armitage, E.~Mele and A.~Vishwanath, Weyl and Dirac semimetals in
  three-dimensional solids, Rev. Mod. Phys \textbf{90},  015001 (2018).

\bibitem{Burkov2015}
A.~A. Burkov, Chiral anomaly and transport in Weyl metals, Journal of Physics:
  Condensed Matter \textbf{27},  113201 (2015).

\bibitem{Bansil2016}
A.~Bansil, H.~Lin and T.~Das, Colloquium: Topological band theory, Rev. Mod.
  Phys \textbf{88},  021004 (2016).

\bibitem{Wan2011}
X.~Wan, A.~M. Turner, A.~Vishwanath and S.~Y. Savrasov, Topological semimetal
  and Fermi-arc surface states in the electronic structure of pyrochlore
  iridates, Phys. Rev. B \textbf{83},  205101 (2011).

\bibitem{Burkov2011a}
A.~A. Burkov, M.~D. Hook and L.~Balents, Topological nodal semimetals, Phys.
  Rev. B \textbf{84},  235126 (2011).

\bibitem{Burkov2011b}
A.~A. Burkov and L.~Balents, Weyl semimetal in a topological insulator
  multilayer, Phys. Rev. Lett. \textbf{107},  127205 (2011).

\bibitem{Hosur2012}
P.~Hosur, S.~A. Parameswaran and A.~Vishwanath, Charge transport in Weyl
  semimetals, Phys. Rev. Lett. \textbf{108},  046602 (2012).

\bibitem{Chang2015}
H.-R. Chang, J.~Zhou, S.-X. Wang, W.-Y. Shan and D.~Xiao, {RKKY} interaction of
  magnetic impurities in Dirac and Weyl semimetals, Phys. Rev. B \textbf{92},
  241103(R) (2015).

\bibitem{Hosseini2015}
M.~V. Hosseini and M.~Askari, Ruderman-Kittel-Kasuya-Yosida interaction in Weyl
  semimetals, Phys. Rev. B \textbf{92},  224435 (2015).

\bibitem{Sun2017}
Y.~Sun and A.~Wang, {RKKY} interaction of magnetic impurities in multi-Weyl
  semimetals, Journal of Physics: Condensed Matter \textbf{29},  435306 (2017).

\bibitem{Yu2010}
R.~Yu, W.~Zhang, H.-J. Zhang, S.-C. Zhang, X.~Dai and Z.~Fang, Quantized
  anomalous hall effect in magnetic topological insulators, Science
  \textbf{329}, ~61 (2010).

\bibitem{Kurebayashi2014}
D.~Kurebayashi and K.~Nomura, Weyl semimetal phase in solid-solution narrow-gap
  semiconductors, Journal of the Physical Society of Japan \textbf{83},  063709
  (2014).

\bibitem{Araki2016}
Y.~Araki and K.~Nomura, Spin textures and spin-wave excitations in doped
  Dirac-Weyl semimetals, Phys. Rev. B \textbf{93},  094438 (2016).

\bibitem{Zhou2018}
J.~Zhou and H.-R. Chang, Dynamical correlation functions and the related
  physical effects in three-dimensional Weyl/Dirac semimetals, Phys. Rev. B
  \textbf{97},  075202 (2018).

\bibitem{Thakur2018}
A.~Thakur, K.~Sadhukhan and A.~Agarwal, Dynamic current-current susceptibility
  in three-dimensional Dirac and Weyl semimetals, Phys. Rev. B \textbf{97},
  035403 (2018).

\bibitem{Ominato2018}
Y.~Ominato and K.~Nomura, Spin susceptibility of three-dimensional Dirac-Weyl
  semimetals, Phys. Rev. B \textbf{97},  245207 (2018).

\bibitem{Ruderman1954}
M.~A. Ruderman and C.~Kittel, Indirect exchange coupling of nuclear magnetic
  moments by conduction electrons, Phys. Rev. \textbf{96}, ~99 (1954).

\bibitem{Kasuya1956}
T.~Kasuya, A theory of metallic ferro- and antiferromagnetism on Zener's model,
  Progress of Theoretical Physics \textbf{16}, ~45 (1956).

\bibitem{Yosida1957}
K.~Yosida, Magnetic properties of Cu-Mn alloys, Phys. Rev. \textbf{106},  893
  (1957).

\bibitem{Kittel1969}
C.~Kittel, \emph{Solid State Physics}, vol.~22, chap. Indirect Exchage
  Interactions in Metals, Academic, New York (1969).

\bibitem{Min2008a}
H.~Min and A.~H. MacDonald, Electronic structure of multilayer graphene,
  Progress of Theoretical Physics Supplement \textbf{176},  227 (2008).

\bibitem{Min2008b}
H.~Min and A.~H. MacDonald, Chiral decomposition in the electronic structure of
  graphene multilayers, PRB \textbf{77},  155416 (2008).

\bibitem{Fang2012}
C.~Fang, M.~J. Gilbert, X.~Dai and B.~A. Bernevig, Multi-weyl topological
  semimetals stabilized by point group symmetry, Phys. Rev. Lett. \textbf{108},
   266802 (2012).

\bibitem{Kobayashi2014}
K.~Kobayashi, T.~Ohtsuki, K.-I. Imura and I.~F. Herbut, Density of states
  scaling at the semimetal to metal transition in three dimensional topological
  insulators, Phys. Rev. Lett. \textbf{112},  016402 (2014).

\bibitem{Sbierski2014}
B.~Sbierski, G.~Pohl, E.~J. Bergholtz and P.~W. Brouwer, Quantum transport of
  disordered Weyl semimetals at the nodal point, Phys. Rev. Lett. \textbf{113},
   026602 (2014).

\bibitem{Pixley2016a}
J.~H. Pixley, D.~A. Huse and S.~Das~Sarma, Rare-region-induced avoided quantum
  criticality in disordered three-dimensional Dirac and Weyl semimetals, PRX
  \textbf{6},  021042 (2016).

\bibitem{Pixley2016b}
J.~H. Pixley, D.~A. Huse and S.~Das~Sarma, Uncovering the hidden quantum
  critical point in disordered massless Dirac and Weyl semimetals, PRB
  \textbf{94},  121107 (2016).

\bibitem{DasSarma2015}
S.~D. Sarma and E.~H. Hwang, Charge transport in gapless electron-hole systems
  with arbitrary band dispersion, Phys. Rev. B \textbf{91},  195104 (2015).

\bibitem{Priour2004}
D.~J. Priour, E.~H. Hwang and S.~D. Sarma, Disordered {RKKY} lattice mean field
  theory for ferromagnetism in diluted magnetic semiconductors, Phys. Rev.
  Lett. \textbf{92},  117201 (2004).

\bibitem{Priour2005}
D.~J. Priour, E.~H. Hwang and S.~D. Sarma, Quasi-two-dimensional diluted
  magnetic semiconductor systems, Phys. Rev. Lett. \textbf{95},  037201 (2005).

\bibitem{Min2017}
H.~Min, E.~H. Hwang and S.~D. Sarma, Ferromagnetism in chiral multilayer
  two-dimensional semimetals, Phys. Rev. B \textbf{95},  155414 (2017).

\bibitem{Ahn2016}
S.~Ahn, E.~H. Hwang and H.~Min, Collective modes in multi-Weyl semimetals,
  Scientific Reports \textbf{6},  34023 (2016).

\bibitem{Kittel2005}
C.~Kittel, \emph{Introduction to Solid State Physics}, Wiley, New Jersey, 8th
  ed. (2005).

\bibitem{DasSarma2003}
S.~D. Sarma, E.~H. Hwang and A.~Kaminski, Temperature-dependent magnetization
  in diluted magnetic semiconductors, Phys. Rev. B \textbf{67},  155201 (2003).

\bibitem{Gradshteyn2015}
I.~S. Gradshteyn and I.~M. Ryzhik, \emph{Tables Of Integrals, Series And
  Products}, Academic, New York, 8th ed. (2015).

\bibitem{AS1964}
M.~Abramowitz and I.~Stegun, \emph{Handbook of Mathematical Functions with
  Formulas, Graphs, and Mathematical Tables}, United States Department of
  Commerce, National Bureau of Standards, Maryland (1964).

\end{thebibliography}

\end{document}